\documentclass{article}
\usepackage[preprint]{spconf}
\usepackage{amsmath,graphicx}

\usepackage{tikz}
\usepackage{standalone}
\usetikzlibrary{positioning, calc, chains, shapes.geometric, shadows, shapes.misc, fit}

\usepackage{multirow,booktabs}
\usepackage{graphicx}

\title{THE VICOMTECH AUDIO DEEPFAKE DETECTION SYSTEM BASED ON WAV2VEC2\\ FOR THE 2022 ADD CHALLENGE}
\name{Juan M. Mart\'in-Do\~{n}as and Aitor \'Alvarez\thanks{This work was supported in part by the Spanish Centre for the Development of Industrial Technology (CDTI) through the Project ÉGIDA—RED DE EXCELENCIA EN TECNOLOGIAS DE SEGURIDAD Y PRIVACIDAD under Grant CER20191012.}}
\address{Vicomtech Foundation, Basque Research and Technology Alliance (BRTA),\\ Mikeletegi 57, 20009 Donostia -- San Sebasti\'an (Spain)\\ \{jmmartin, aalvarez\}@vicomtech.org}
\begin{document}
%
\maketitle
\begin{abstract}
This paper describes our submitted systems to the 2022 ADD challenge withing the tracks 1 and 2. Our approach is based on the combination of a pre-trained wav2vec2 feature extractor and a downstream classifier to detect spoofed audio. This method exploits the contextualized speech representations at the different transformer layers to fully capture discriminative information. Furthermore, the classification model is adapted to the application scenario using different data augmentation techniques. We evaluate our system for audio synthesis detection in both the ASVspoof 2021 and the 2022 ADD challenges, showing its robustness and good performance in realistic challenging environments such as telephonic and audio codec systems, noisy audio, and partial deepfakes.
\end{abstract}
\begin{keywords}
antispoofing, wav2vec2, audio deepfakes, self-supervised, data augmentation
\end{keywords}
\section{Introduction}
\label{sec:intro}

Speech synthesis and voice conversion technologies \cite{sisman21} have rapidly grown in the last years, mainly thanks to the development of the deep learning paradigm. Although this broadens the application of these technologies, a threat is also present: the generation of deepfake speech that can even foolish modern automatic speaker verification systems \cite{tan21}. The need for reliable countermeasures has favored the research on audio deepfake detection systems able to detect spoofed speech \cite{wu15}. An example of this effort is the ASVspoof series \cite{wu15asv,todisco19}, consisting of biannual challenges focused on the development of antispoofing countermeasures for verification systems.

The continuous improvements in audio deepfake detection have widespread interest in developing robust solutions in realistic challenging scenarios. For example, the systems robustness on noisy and reverberant scenarios was studied in \cite{tian16,gomez19}. Likewise, the ASVspoof2021 challenge \cite{yamagishi21} considered synthesized speech transmitted over different telephone networks, as well as speech deepfakes modified through commercial compression algorithms. Moreover, recent works \cite{zhang21} have explored the detection of partially spoofed audio. In order to continue improving these systems in complex environments, the 2022 Audio Deep synthesis Detection (ADD) challenge has been recently launched \cite{yi22}. Its main goal is the detection of deep synthesis and manipulated audios, and it includes three different tracks: 1) Low-quality fake audio detection (diverse background noises and disturbances are contained in the audios); 2) Partially fake audio detection (several small fake speech segments are hidden in real audio); and 3) Audio fake game.

This paper presents our contributions to the ADD 2022 challenge, based on the use of the wav2vec2 (W2V2) \cite{baevski20} approach to extract discriminative information that helps detect spoofed audio. These models are trained with self-supervised learning methods with a large amount of unlabelled speech data, allowing to learn high-level representations of the speech signal. W2V2 has been explored in different speech processing tasks, such as speech recognition, speaker verification \cite{chen21} and emotion classification \cite{pepino21}. Regarding audio deepfake detection, only a few works have explored this approach or similar \cite{xie21,wang21}. Differently from previous works, we propose to use a pre-trained W2V2 model as a feature extractor, but using the encoded representations of the different transformer layers. This information can be then exploited by a simple but effective downstream model to detect spoofing attacks. Moreover, we analyzed different data augmentation techniques to adapt the classifier to the final application scenario. Our approach was evaluated in both tracks 1 and 2 of the ADD challenge, where it achieved first and fourth position among the participants, respectively. In addition, we include results on the logical access and speech deepfake tracks of the ASVspoof 2021.

The remainder of this paper is organized as follows. In Section~\ref{sec:proposal}, we present our proposal based on the W2V2 approach. Section~\ref{sec:framework} describes the speech databases used for the evaluations and the data augmentation training strategies. Then, in Section~\ref{sec:res}, the results are presented and analyzed. Finally, conclusions are summarized in Section~\ref{sec:conclusions}.

\section{wav2vec2-based audio deepfake detection approach}
\label{sec:proposal}

\begin{figure}[!t]
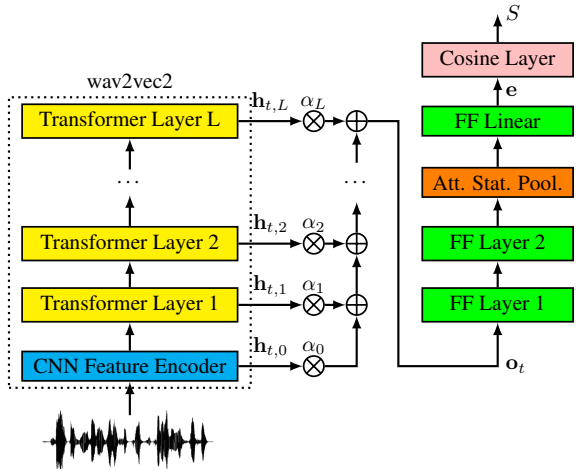

  \centering
  \includestandalone[width=0.9\linewidth,trim={0 1.2em 0 0},clip]{scheme}
  \caption{Overview of the proposed audio deepfake detection approach based on wav2vec2.}
  \label{fig:w2v2}
\end{figure}

A diagram of our proposed system for the ADD challenge is depicted in \figurename{~\ref{fig:w2v2}}. It is composed of a W2V2 feature extractor, which obtains the encoded speech representations, and a classification model that scores the input audio as genuine or spoof. A detailed explanation of each part is presented in the next subsections.

\subsection{Wav2vec2 as Feature Extractor}

As the base W2V2 architecture, we evaluated the Large models of 300M parameters, presented in \cite{babu21}, trained with 53 and 128 languages (XLS-53 and XLS-128, respectively). The raw speech signal is first processed by a feature encoder composed of several convolutional layers (CNN), which extracts vector representations of size 1024 every 20 ms, using a receptive field of 25 ms. Then, these encoder features are fed into a transformer network with 24 layers which is used to obtain contextualized representations of the speech signal. The model is trained in a self-supervised setting using a contrastive loss. The objective is to predict the quantized representations of certain masked encoded feature representations from a set of distractors using the contextualized representations. Thus, this model can learn high-level representations of the speech signal from a large amount of unlabelled data. The features extracted from the pre-trained W2V2 model can be used to train a downstream classifier in a specific speech processing task with a relatively few amount of labeled data. Moreover, the W2V2 and downstream models can be jointly trained in the related task. In this work, we explore the use of W2V2 as a pre-trained model, thus allowing to have a general feature extractor that can be used with specialized classification models adapted to each spoofing detection task.

\subsection{Classification Model}

\begin{table}[!t] 
\caption{Architecture of the downstream classification model. It includes each layer and its output dimension, where $T$ is the number of time frames and $N$ the size of the mini-batch.}
\label{table:classmodel}
\centering
\vspace{1em}
\resizebox{0.9\linewidth}{!}{
\begin{tabular}{c|c}
\hline
\textbf{Layer name}         & \textbf{Output size} \\ \hline
W2V2 features               & $N$ $\times$ T $\times$ 1024 $\times$ 25                    \\
Temp. Norm. + Layer weight. & $N$ $\times$ T $\times$ 1024                    \\
FF Layer (1 and 2)          & $N$ $\times$ T $\times$ 128                    \\
Att. Stat. Pool.            & $N$ $\times$ 256                   \\
FF Linear                   & $N$ $\times$ 128                    \\
Cosine Layer                & $N$                    \\ \hline
\end{tabular}
}
\end{table}

The contextualized representations from the last transformer layer of a pre-trained model can be useful for certain speech tasks, like speech recognition. Nevertheless, previous works have shown that for other tasks, for example, speaker verification or emotion recognition, more discriminative information can be obtained from the first or intermediate layers \cite{chen21,pepino21}. Therefore, we followed the methodology of these previous works and used the hidden representations of the different transformer layers as input for our downstream model.

The classification model processes the W2V2 hidden representations as follows. First, a temporal normalization \cite{ulyanov16} is applied at the input features from each transformer layer. Then, for each temporal step $t$, an output representation is computed from the normalized hidden representations $\mathbf{h}_{t,l}$ as $\mathbf{o}_{t} = \sum_{l=0}^{L} \alpha_{l} \mathbf{h}_{t,l}$,
where $l$ represents the hidden layer index, $L$ the number of transformer layers, and $\alpha_{l}$ are network trainable weights, which are normalized to sum one \cite{pepino21}. The computed vectors at each time step are fed into two feed-forward (FF) layers with ReLU activation and dropout, and then they are passed to an attentive statistical pooling layer \cite{okabe18}. This attention layer computes and concatenates the temporal mean and standard deviation of the vectors at the different time steps, thus obtaining a single representation for the whole utterance. A linear layer (FF without non-linearity) is then applied to compute the embedding vector $\mathbf{e}$. Finally, the final score is obtained as a cosine similarity, $S = \cos(\mathbf{w}, \mathbf{e}) \in [-1,1]$, where $\mathbf{w}$ is a vector network parameter representing the direction of genuine speech in the embedding space. The classification model is shown in more detail in Table~\ref{table:classmodel}. The model is trained to compute higher scores for genuine speech using a One-class softmax loss function \cite{zhang21spl}.



\section{Experimental framework}
\label{sec:framework}

This section describes the speech databases used to train and evaluate our systems, as well as the data augmentation techniques and training setup procedure. For each of the challenges, we trained our systems using only the corresponding train data of that challenge (closed conditions).

\subsection{ADD 2022 challenge database}

The ADD 2022 database \cite{yi22} comprises spoofed audio generated by speech synthesis and voice conversion systems. The train and development (dev) sets contain clean speech based on the multi-speaker Mandarin speech corpus AISHELL-3 \cite{shi20}. Each set has about 28K utterances, with different speakers each. Tracks 1 and 2 include both an adaptation set with about 1K utterances, and a test set with about 100K utterances without labels. The adaptation set presents similar conditions that the audios in the test set, and it is provided to adapt the systems trained with the general train set. For track 1, the corresponding sets incorporate both genuine and spoof utterances with various real-world noises and background music. For track 2, the corresponding sets include genuine utterances and partially fake utterances generated by manipulating the original genuine speech with real or synthesized audio.

\subsection{ASVspoof 2021 challenge database}

We focused on the logical access (LA) and speech deepfake (DF) partitions of the ASVspoof 2021 challenge \cite{yamagishi21}. Both partitions contain spoofed audio generated by speech synthesis or voice conversion methods, with clean speech derived from the VCTK corpus \cite{yamagishi12}. The 2021 challenge only considered evaluation data, so the ASVspoof 2019 \cite{todisco19} LA train and dev sets were used for training. Each set contains about 25K utterances from 20 and 10 speakers, respectively. The 2021 LA and DF sets are similar to the 2019 LA data, but they consider more challenging scenarios. The LA set contains about 180K utterances of speech transmitted through real telephonic systems at different bandwidths and with different codecs. The DF set includes about 600K utterances of speech processed with various commercial audio codecs. Moreover, this set considers clean speech from other databases, making it more challenging.

\subsection{Data augmentation techniques}
\label{subsec:data_aug}

In order to improve the performance of our approach and to adapt it to the different scenarios, we considered the use of data augmentation techniques to be applied \textit{on the fly} during the training step. The main augmentation technique evaluated corresponded to the use of low-pass finite impulse response (FIR) filters on the speech signals. This procedure has shown its success for ASVspoof 2021 LA and DF tracks \cite{tomilov21} as it emulates the effects of transmission artifacts and codecs. Moreover, they act by masking frequencies in the speech signal, improving the generalization capabilities of LA detection systems. Thus, they can be also useful for the conditions presented in the ADD database. In addition, they are directly applied to the raw waveform, making them compatible with the W2V2 network. We evaluated both narrowband (NB) and wideband (WB) FIR filters, following a similar procedure to that presented in \cite{tomilov21}.

For the ADD challenge, we joined the train and corresponding adaptation set of each track to improve the model generalization on the new challenging conditions. Although they are scarce, the adaptation data can be helpful for the W2V2-based system to be adapted to the corresponding scenario. Furthermore, for track 2 we also considered the generation of new partially fake audios. To this end, 20\% of the genuine utterances in the train/dev set were randomly selected at each epoch. For each of these utterances, we selected a segment of variable duration, shorter than the original audio, from a different utterance (genuine or spoof) within the corresponding set. The segment was then overlapped over the original utterance in a random position.

\subsection{Training setup}

The models were trained using the Adam optimizer \cite{kingma15} with the default learning rate and a dropout of 0.2. During training, the W2V2 parameters were frozen, and only the classifier parameters were updated. A mini-batch of 8 utterances was used, and the parameters were updated every 8 iterations. At each epoch, the model was evaluated using the corresponding dev set, keeping the model with the lowest loss. The training stopped after 10 epochs without improvements on the dev set.

\section{Results}
\label{sec:res}

In this section, we describe the results obtained for our proposal and its different configurations in both ASVspoof 2021 and 2022 ADD challenges. The different systems were evaluated and compared in terms of equal error rate (EER), which is the most common metric used in biometric applications.

\subsection{Results on ADD challenge}

Table~\ref{table:add_res} shows the results obtained for tracks 1 and 2 of the ADD challenge. Our approach is evaluated in terms of the W2V2 model and data augmentation techniques, as well as the ADD data sets used for training. The XLS-128 model shows better results in general, but the main improvements come from the use of adaptation data along with the train set. The use of a small portion of data with similar conditions to the test set helps the model to adapt better to the scenario evaluated, reducing EER drastically. Moreover, in track 2 we generated more adaptation data through the partial fake augmentation strategy described in Subsection~\ref{subsec:data_aug}, increasing even further the model discrimination capabilities. Finally, we explored the use of NB FIR filters to increase the robustness under low-quality and partial fake conditions. This strategy reduced the EER by about 1\% in both tracks and combined with the previous ones, it achieved the best results. Thus, this demonstrates that the FIR-based strategy is not only useful to emulate telephonic or codec conditions \cite{tomilov21}, but also helps to train better antispoofing systems in challenging conditions.

\begin{table}[!t] 
\caption{Final results in terms of EER (\%) of our submitted approaches to the ADD 2022 challenge. It also includes different variants for the W2V2 model and the data augmentation (DA) and adaptation strategies.}
\label{table:add_res}
\centering
\vspace{1em}
\resizebox{0.9\linewidth}{!}{
\begin{tabular}{lllcc}
\toprule
\textbf{W2V2} & \textbf{Sets} & \textbf{DA} & \textbf{Track1} & \textbf{Track2} \\
\midrule
\multirow{2}*{XLS-53} & Train & -                 &   32.96  &  38.09  \\
& Tr.+Adap. & - & 23.70 & 33.73 \\
\midrule
\multirow{5}*{XLS-128} & Train & -                 &   32.20  &  45.88  \\
& Tr.+Adap. & - & 22.62 & 30.35 \\
& Tr.+Adap. & FIR & \textbf{21.71} & - \\
& Tr.+Adap. & partial & - & 17.58 \\
& Tr.+Adap. & FIR+part. & - & \textbf{16.59} \\
\bottomrule
\end{tabular}
}
\end{table}

\subsection{Results on ASVspoof 2021}

\begin{table}[!t] 
\caption{EER (\%) results of our proposed W2V2 approach on ASVspoof 2021, considering different pre-trained models and FIR-based data augmentation strategies.}
\label{table:asv21_ours}
\centering
\vspace{1em}
\resizebox{0.9\linewidth}{!}{
\begin{tabular}{llcc}
\toprule
\textbf{W2V2 model} & \textbf{Data augmentation} & \textbf{LA} & \textbf{DF} \\
\midrule
\multirow{3}*{XLS-53} & -                 &   8.87  &  7.71  \\
& FIR-NB & 4.34 & 11.27 \\
& FIR-WB & 4.98 & 6.99 \\
\midrule
\multirow{3}*{XLS-128} & -                 &   7.20  &  5.68  \\
& FIR-NB & \textbf{3.54} & 6.18 \\
& FIR-WB & 7.08 & \textbf{4.98} \\
\bottomrule
\end{tabular}
}
\end{table}

Table~\ref{table:asv21_ours} shows the results obtained for our approach in the LA and DF tracks of the ASVspoof21 database. We experimented with different combinations of W2V2 pre-trained models and low-pass FIR-based data augmentations, both NB and WB filters. Again, the XLS-128 model performs the best in both partitions. Furthermore, the FIR-based augmentations help to improve the discrimination capabilities of our proposal. Particularly, the NB filters are useful for LA scenarios as they emulate traditional telephonic systems, while the DF is more favored by the WB filters. WB filters emulate general audio codecs as the ones in the DF set.

In addition, in Table~\ref{table:asv21_comp} we compare our approach with other systems in this challenge. As it is shown, our approach outperforms other methods in the DF set and achieves competitive results in LA track, despite being a single system (not an ensemble of classifiers). The W2V2 features show robustness to the varied speech content in the DF set, allowing to highly reduce EER compared to other systems. On the other hand, compared with \cite{wang21}, our approach exploits the representations of the different transformer layers, allowing comparative results while using a simpler downstream model. Moreover, the classifier can be effectively adapted using data augmentations techniques, while the W2V2 pre-trained model remains as a general feature extractor. It can be interesting in practical applications to save computational resources.

\begin{figure}[!t]
  \centering
  \includegraphics[width=\linewidth]{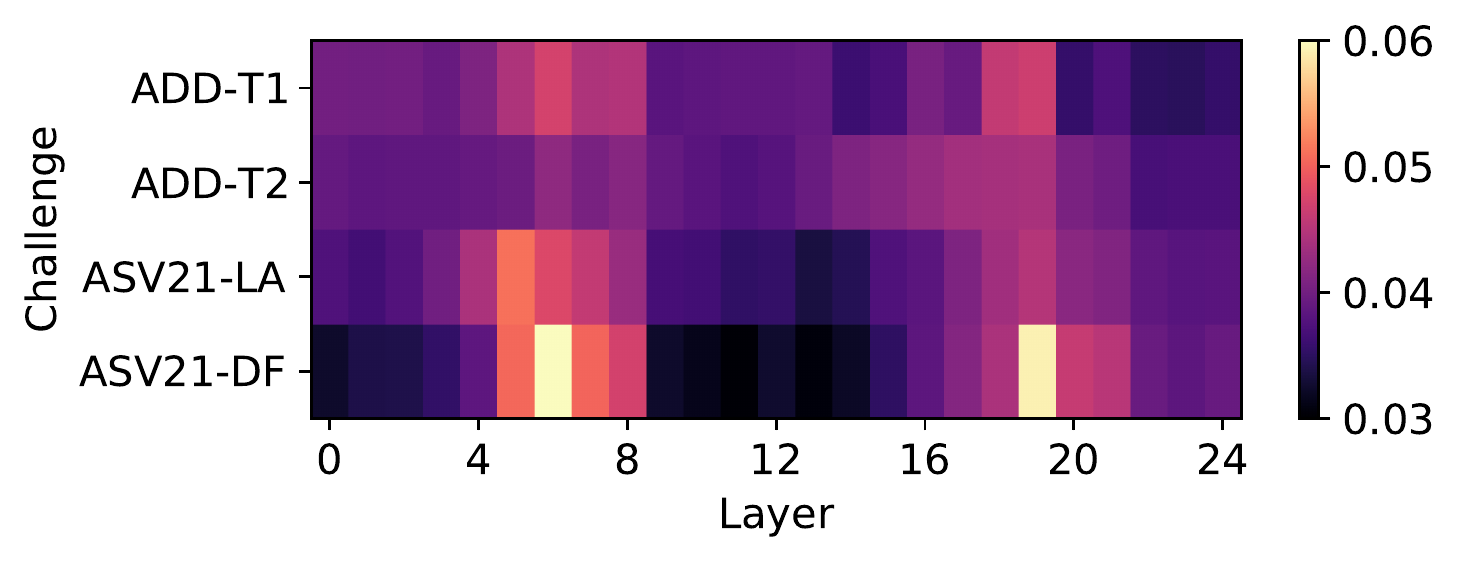}
  \caption{Visualization of the weight values $\alpha_l$ for our best-proposed approach at each challenge and track.}
  \label{fig:alpha}
\end{figure}

Finally, \figurename{~\ref{fig:alpha}} draws the value of the weights $\alpha_l$ for our best systems in the tracks of ADD 2022 and ASVspoof21 challenges. As it can be observed, the classifier uses the information from the different transformer layers to detect spoofed audio, using different layer weights depending on the scenario considered.

\begin{table}[!t] 
\caption{Comparative EER (\%) results of our proposed method with participant systems in the ASVspoof 2021 challenge and other self-supervised approaches.}
\label{table:asv21_comp}
\centering
\vspace{1em}
\resizebox{0.9\linewidth}{!}{
\begin{tabular}{lcc}
\toprule
\textbf{System} & \textbf{LA} & \textbf{DF} \\
\midrule
LCNN+ResNet+RawNet \cite{tomilov21} & \textbf{1.32} & 15.64 \\
GMM+LCNN (Ensemble) \cite{das21} & 3.62 & 18.30 \\
ECAPA-TDNN (Ensemble) \cite{chen21asv} & 5.46 & 20.33 \\
ResNet (Ensemble) \cite{chen21pin} & 3.21 & 16.05 \\
W2V2 (fixed)+LCNN+BLSTM \cite{wang21} & 10.97 & 7.14 \\
W2V2 (finetuned)+LCNN+BLSTM \cite{wang21} & 7.18 & 5.44 \\
\midrule
\textit{Proposed system} & 3.54 & \textbf{4.98} \\
\bottomrule
\end{tabular}
}
\end{table}

\section{Conclusion}
\label{sec:conclusions}

In this work, we have presented our proposed system for the 2022 ADD challenge, which is based on the pre-trained W2V2 self-supervised architecture. Our approach exploits the contextualized representations at the different transformer layers in a finetuned classification model to detect spoofed speech. While the W2V2 model remains general, we adapted the downstream network to each scenario through data augmentation techniques and adaptation data. Our proposed method shows competitive results in both the ASVspoof 2021 and 2022 ADD challenges, which represent challenging realistic scenarios with synthesized spoofed audio. Thus, our system ranked first and fourth position in tracks 1 and 2 of 2022 ADD challenge, respectively. As future work, we will test other self-supervised models as well as additional data augmentation techniques.

\vfill\pagebreak

\bibliographystyle{IEEEbib}
\bibliography{refs}

\begin{thebibliography}{10}

\bibitem{sisman21}
B.~Sisman, J.~Yamagishi, S.~King, and H.~Li,
\newblock ``{An overview of voice conversion and its challenges: From
  statistical modeling to deep learning},''
\newblock {\em IEEE/ACM Trans. on Audio, Speech, and Lang. Processing}, vol.
  29, pp. 132--157, 2021.

\bibitem{tan21}
C.~B. Tan et~al.,
\newblock ``{A survey on presentation attack detection for automatic speaker
  verification systems: State-of-the-art, taxonomy, issues and future
  direction},''
\newblock {\em Multimedia Tools and Applications}, vol. 80, no. 21, pp.
  32725--32762, 2021.

\bibitem{wu15}
Z.~Wu, N.~Evans, T.~Kinnunen, J.~Yamagishi, F.~Alegre, and H.~Li,
\newblock ``Spoofing and countermeasures for speaker verification: {A
  survey},''
\newblock {\em Speech Communication}, vol. 66, pp. 130--153, 2015.

\bibitem{wu15asv}
Z.~Wu et~al.,
\newblock ``{ASVspoof 2015: The first automatic speaker verification spoofing
  and countermeasures challenge},''
\newblock in {\em Proc. Interspeech 2015}, 2015, pp. 2037--2041.

\bibitem{todisco19}
M.~Todisco et~al.,
\newblock ``{ASVspoof 2019: Future horizons in spoofed and fake audio
  detection},''
\newblock in {\em Proc. Interspeech 2019}, 2019, pp. 1008--1012.

\bibitem{tian16}
X.~Tian, Z.~Wu, X.~Xiao, E.~S. Chng, and H.~Li,
\newblock ``An investigation of spoofing speech detection under additive noise
  and reverberant conditions.,''
\newblock in {\em Proc. Interspeech 2016}, 2016, pp. 1715--1719.

\bibitem{gomez19}
A.~Gomez-Alanis, A.~M. Peinado, J.~A. Gonzalez, and A.~M. Gomez,
\newblock ``A gated recurrent convolutional neural network for robust spoofing
  detection,''
\newblock {\em IEEE/ACM Trans. on Audio, Speech, and Lang. Processing}, vol.
  27, no. 12, pp. 1985--1999, 2019.

\bibitem{yamagishi21}
J.~Yamagishi et~al.,
\newblock ``{ASVspoof 2021: Accelerating progress in spoofed and deepfake
  speech detection},''
\newblock in {\em Proc. 2021 ASVspoof Workshop}, 2021, pp. 47--54.

\bibitem{zhang21}
L.~Zhang, X.~Wang, E.~Cooper, J.~Yamagishi, J.~Patino, and N.~Evans,
\newblock ``An initial investigation for detecting partially spoofed audio,''
\newblock in {\em Proc. Interspeech 2021}, 2021, pp. 4264--4268.

\bibitem{yi22}
J.~Yi et~al.,
\newblock ``{ADD 2022: The} first audio deep synthesis detection challenge,''
\newblock in {\em Proc. ICASSP}, 2022.

\bibitem{baevski20}
A.~Baevski, A.~Zhou, H.and~Mohamed, and M.~Auli,
\newblock ``wav2vec 2.0: A framework for self-supervised learning of speech
  representations,''
\newblock {\em arXiv preprint arXiv:2006.11477}, 2020.

\bibitem{chen21}
Z.~Chen, S.~Chen, Y.~Wu, Y.~Qian, C.~Wang, S.~Liu, Y.~Qian, and M.~Zeng,
\newblock ``Large-scale self-supervised speech representation learning for
  automatic speaker verification,''
\newblock {\em arXiv preprint arXiv:2110.05777}, 2021.

\bibitem{pepino21}
L.~Pepino, P.~Riera, and L.~Ferrer,
\newblock ``{Emotion Recognition from Speech Using wav2vec 2.0 Embeddings},''
\newblock in {\em Proc. Interspeech 2021}, 2021, pp. 3400--3404.

\bibitem{xie21}
Y.~Xie, Z.~Zhang, and Y.~Yang,
\newblock ``Siamese network with wav2vec feature for spoofing speech
  detection,''
\newblock in {\em Proc. Interspeech}, 2021, pp. 4269--4273.

\bibitem{wang21}
X.~Wang and J.~Yamagishi,
\newblock ``Investigating self-supervised front ends for speech spoofing
  countermeasures,''
\newblock {\em arXiv preprint arXiv:2111.07725}, 2021.

\bibitem{babu21}
A.~Babu et~al.,
\newblock ``{XLS-R: Self-supervised} cross-lingual speech representation
  learning at scale,''
\newblock {\em arXiv preprint arXiv:2111.09296}, 2021.

\bibitem{ulyanov16}
D.~Ulyanov, A.~Vedaldi, and V.~Lempitsky,
\newblock ``{Instance normalization: The missing ingredient for fast
  stylization},''
\newblock {\em arXiv preprint arXiv:1607.08022}, 2016.

\bibitem{okabe18}
K.~Okabe, T.~Koshinaka, and K.~Shinoda,
\newblock ``Attentive statistics pooling for deep speaker embedding,''
\newblock in {\em Proc. Interspeech 2018}, 2018, pp. 2252--2256.

\bibitem{zhang21spl}
Y.~Zhang, F.~Jiang, and Z.~Duan,
\newblock ``One-class learning towards synthetic voice spoofing detection,''
\newblock {\em IEEE Signal Processing Letters}, vol. 28, pp. 937--941, 2021.

\bibitem{shi20}
H.~Shi, Y.and~Bu, X.~Xu, S.~Zhang, and M.~Li,
\newblock ``{AISHELLl-3: A multi-speaker mandarin TTS corpus and the
  baselines},''
\newblock {\em arXiv preprint arXiv:2010.11567}, 2020.

\bibitem{yamagishi12}
J.~Yamagishi,
\newblock ``English multi-speaker corpus for {CSTR} voice cloning toolkit,''
  2012.

\bibitem{tomilov21}
A.~Tomilov, A.~Svishchev, M.~Volkova, A.~Chirkovskiy, A.~Kondratev, and
  G.~Lavrentyeva,
\newblock ``{STC Antispoofing Systems for the ASVspoof2021 Challenge},''
\newblock in {\em Proc. 2021 ASVspoof Workshop}, 2021, pp. 61--67.

\bibitem{kingma15}
D.~P Kingma and J.~Ba,
\newblock ``{Adam: A method for stochastic optimization},''
\newblock in {\em Proc. ICLR}, 2015.

\bibitem{das21}
R.~K. Das,
\newblock ``{Known-unknown Data Augmentation Strategies for Detection of
  Logical Access, Physical Access and Speech Deepfake Attacks: ASVspoof
  2021},''
\newblock in {\em Proc. 2021 ASVspoof Workshop}, 2021, pp. 29--36.

\bibitem{chen21asv}
X.~Chen, Y.~Zhang, G.~Zhu, and Z.~Duan,
\newblock ``{UR Channel-Robust Synthetic Speech Detection System for ASVspoof
  2021},''
\newblock in {\em Proc. 2021 ASVspoof Workshop}, 2021, pp. 75--82.

\bibitem{chen21pin}
T.~Chen, E.~Khoury, K.~Phatak, and G.~Sivaraman,
\newblock ``{Pindrop Labs’ Submission to the ASVspoof 2021 Challenge},''
\newblock in {\em Proc. 2021 ASVspoof Workshop}, 2021, pp. 89--93.

\end{thebibliography}

\end{document}